\documentstyle[12pt]{article}

\makeatletter
\@addtoreset{equation}{section}

\makeatother

\begin{document}

\newcommand{\tr}{\bigtriangleup}
\newcommand{\vp}[2]{\frac{\,\delta #1}{\delta #2 }}
\newcommand{\vpi}[2]{\frac{\,\delta #1}{i\delta #2 }}
\newcommand{\imb}[1]{\!\not{\!#1}}
\newcommand{\con}{<\!\!{\bar \psi}\psi\!\!>}

\author{V.E.~Rochev\thanks{rochev@mx.ihep.su}\hspace{0.5em}
and P.A.~Saponov\thanks{saponov@mx.ihep.su}\\
{\footnotesize \it 
Institute for High Energy Physics, 142284 Protvino, Moscow 
region, Russia}}
\title{The four-fermion interaction in D=2,3,4:\\
a nonperturbative treatment.}
\date{}

\maketitle

\begin{abstract}
A new nonperturbative approach is used to investigate the
Gross-Neveu model of four fermion interaction in the
space-time dimensions 2, 3 and 4, the number $N$ of inner
degrees of freedom being a fixed integer. The spontaneous
symmetry breaking is shown to exist in $D=2,3$ and the
running coupling constant is calculated. The four
dimensional theory  seems to be trivial.
\end{abstract}

\section{Introduction}

In spite of  its great and numerous successes the
perturbation theory cannot describe a wide class of
important phenomena (like confinement or spontaneous
symmetry breaking)  playing the key role in the problem of
full and comprehensive description of physical reality.  So
it is quite natural that considerable efforts are applied
in order to develop the nonperturbative methods in the
quantum field theory.  So far, only a few sufficiently
effective methods like the effective potential, the $1/N$
expansion, the  Gauss effective potential method
\cite{Stiv} or the variational perturbation theory
\cite{Solo} are known.

A new nonperturbative approach has been recently proposed
in work \cite{Roch1}. The ability of the method was
demonstrated by the example of a self interacting scalar
field in various dimensions.  In the present paper we would
like to investigate the Gross-Neveu model of the fermion
fields with an arbitrary {\it fixed} number $N$ of inner
degrees of freedom. The cases  $D=2,3,4$ were elaborated
and the spontaneous symmetry breaking was found to exist in
two and three dimensions. For the four dimensional
Gross-Neveu model our consideration gives the arguments in
favor of the triviality of this model. These results
exhibit the efficiency of the method and they are the
finite $N$ generalization of the known results  obtained in
the framework of $1/N$ expansion \cite{GrNe}--\cite{Kim}.

The paper is organized as follows. Section~2 consists of a
short introduction to the method of \cite{Roch1} to be used
throughout the paper. Sections~3, 4 and 5 are devoted to
the Gross-Neveu model in the two, three and four
dimensions, respectively. Section~6 contains another
approach to the subject based on the bilocal source
formalism.

\section{The method}
\label{Method}

In our approach we will use one of the most suitable tools
for the nonperturbative treatment of a quantum field model
--- the Schwinger-Dyson equation. As the system of
Schwinger-Dyson equations for the Green functions consists
of the infinite number of mutually connected equations, one
should truncate it in some way in order to find  an
approximate solution.  It is obvious enough that the
concrete way of truncation has a crucial significance for
the results. For example, if we solve the Schwinger-Dyson
system iteratively by expanding the Green functions into
the series in the coupling constant, we obtain the
perturbative solution. Being the simplest from the
practical point of view the perturbative approach is the
worst in the mathematical sense. The matter is that a small
parameter (the coupling constant) is a multiplier at the
highest derivative term of the functional differential
Schwinger-Dyson  equation for the Green functions
generating functional.  This means that the equation is a
singularly perturbed one.  So, the perturbative procedure
is valid only for the restrictive class of boundary
conditions and cannot, in principle catch all solutions
\cite{RochSap1}.

Taking into account the reasons above, we can conclude that
a proper approximation scheme for the nonperturbative
solution of the Schwinger-Dyson equation should obey the
following requirements:

\begin{itemize}
\item
It must take into account the highest derivative term of
the Schwinger-Dyson equation already at the leading
approximation.
\item
It must allow one to make the renormalization procedure.
\item
It should be simple enough for the practical calculations.
\end{itemize}

An approximation scheme proposed in \cite{Roch1} obeys the
requirements listed above. Below we give a short
introduction to the main ideas of the method for the
reader's convenience.

Consider the  theory of a self-interacting scalar field
$\phi(x)$ with the action

\begin{equation}
S(\phi) = \int dx \left(\frac{1}{2} (\partial_{\mu}\phi)^2
 -\frac{\mu^2}{2} \phi^2 - \lambda\,\phi^4\right).
\end{equation}
The generating functional of  $n$-point Green functions can
be written as follows

\begin{equation}
G = \sum_{n=0}^{\infty} {G_{n} j^n}, 
\end{equation}
where $j(x)$ is a field source. The $n$-th derivative of G
at $j=0$  is the $n$-point Green function $G_{n}$.

The Schwinger-Dyson equation for the generating functional
of this model reads

\begin{equation}
(\mu^2 + \partial^2) \frac{\delta G}{\delta j(x)} +
4\lambda \frac{\delta^3 G}{\delta j^3(x)} 
-i j(x) G  = 0\;.
\end{equation}

The central idea of the iterative scheme is to consider the
{\it last} term of this equation as a perturbation to the
leading approximation. That is we take the following
"equation with constant coefficients" as the leading
approximation:

\begin{equation}
(\mu^2 + \partial^2) \frac{\delta G^{(0)}}{\delta j(x)} +
4\lambda \frac{\delta^3 G^{(0)}}{\delta j^3(x)} 
 = 0\;.\label{ledapp}
\end{equation}
Presenting the full functional $G(j)$ as the sum

\begin{equation}
G(j) = \sum_n G^{(n)}(j) \label{genfun}
\end{equation}
we then write for the terms of this sum the recursive chain
of equations:

\begin{equation}
(\mu^2 + \partial^2) \frac{\delta G^{(n)}}{\delta j(x)} +
4\lambda \frac{\delta^3 G^{(n)}}{\delta j^3(x)} 
= i j(x) G^{(n-1)}  \;.\label{rec}
\end{equation}

The solution for (\ref{ledapp}) is sought for in the form
$G^{(0)} = \exp{(i\sigma j)}$ and from (\ref{ledapp}) one
obtains a characteristic equation for the function
$\sigma(x)$. For the $G^{(n)}$ we put $G^{(n)} =
P_n(j)G^{(0)}$, where $P_n(j)$ is a polynomial in $j(x)$
with unknown coefficients to be defined from (\ref{rec}).
These coefficients define the Green  functions of the
corresponding step of the iteration scheme. It should be
noted, that at the leading approximation one defines "the
vacuum" of the model, at the first step the connected part
of the propagator enters the game, whereas for the higher
Green functions one can define an approximant for the
disconnected part only.  At the next steps of the scheme we
calculate many-particle amplitudes and corrections to the
propagator and so on.

This procedure is  model independent and has a regular
character.  The last property is due to the generating
functional is regular at $j=0$ {\it by definition}, as its
derivatives at this point  are the Green functions of the
model. Therefore the perturbation theory around the point
$j=0$ is {\it regular} recipe for solving the
Schwinger-Dyson equation.

The renormalization procedure can be easily introduced into
the scheme (see \cite{Roch1} for details). Shortly the
renormalization can be carried out in the following few
steps.

{\bf (i)}
 All the necessary counter terms are expanded in a sum,
analogous to  (\ref{genfun}):

\begin{equation}
\mu^2\rightarrow \mu^2 + \delta\mu^2_{(0)} +
\delta\mu^2_{(1)} +\, \dots
\qquad \lambda \rightarrow \lambda + \delta\lambda_{(0)} +
\delta\lambda_{(1)} +\, \dots  \quad {\rm etc.},
\end{equation}
where the subscript of a counter term stands for the
approximation scheme step at which this counter term should
be taken into account.
 
{\bf (ii)} Equations (\ref{ledapp}) and (\ref{rec}) are
modified respectively:

\begin{equation}
(\mu^2 +\delta\mu^2_{(0)} + (1+\delta Z^{(0)}_{\phi})\,
\partial^2) \frac{\delta G^{(0)}}{\delta j(x)} +
4(\lambda +\delta\lambda_{(0)}) 
\frac{\delta^3 G^{(0)}}{\delta j^3(x)} 
 = 0
\end{equation}

\begin{eqnarray}
(\mu^2 +\delta\mu^2_{(0)} + (1+\delta Z^{(0)}_{\phi})\,
\partial^2) 
\frac{\delta G^{(1)}}{\delta j(x)} &+&
4(\lambda +\delta\lambda_{(0)}) 
\frac{\delta^3 G^{(1)}}{\delta j^3(x)} 
\nonumber\\
&=& i j(x) G^{(0)} -  \delta\mu^2_{(1)}
\frac{\delta G^{(0)}}{\delta j(x)} 
-\delta Z^{(1)}_{\phi}\partial^2 
\frac{\delta G^{(0)}}{\delta j(x)} 
\end{eqnarray}
and so on.

{\bf (iii)}
The counter terms entering the scheme at the $n$-th step
can be fixed only at the next step. Before the
renormalization of the next step has been done, the
equations of the $n$-th step are  nothing but some
relations among the counter terms.

Now we are ready to go to the subject of our paper --- the
Gross-Neveu model of the four fermion interaction
\cite{GrNe}.

\section{The general consideration and D=2 case}
\label{d2}

In \cite{GrNe} D.~Gross and A.~Neveu  have investigated the
model  of the $N$-coloured spinor fields with scalar-scalar
four fermion interaction at large $N$. Most of the works
devoted to the Gross-Neveu model also deals with the $1/N$
expansion \cite{ZinnJ}--\cite{Rosen}. Our goal is to
investigate the Gross-Neveu model when the number of
colours $N$ is an arbitrary fixed integer.

The model is defined in the $D$-dimensional Minkowski
space-time by the action:

\begin{equation}
S =\int\,d^Dx \left({\bar \psi}\,(i\!\not{\!\partial} - m)\,
\psi + \frac{\lambda}{2}\,({\bar \psi}\psi)^2\right),
\label{act}
\end{equation}
where $\psi_k(x)$ is a spinor field with $N$ isotopic
degrees of freedom.  The summation over the isotopic
indices  $k$ is implicit in (\ref{act}).  The mass
dimension of the coupling constant $\lambda$ is zero in the
two dimensional space-time, therefore  model (\ref{act}) is
renormalizable in the two dimensions.

Let us transform our model to that of the  spinor and
scalar fields coupled by Yukawa interaction \cite{GrNe}:

\begin{equation}
S_{\mbox{\scriptsize eff}} 
= \int\,d^Dx \left({\bar \psi}\,(i\!\not{\!\partial} - m)\,
\psi - {1\over 2}\phi(\mu^2 + \partial^2)\phi +
\beta\,\phi({\bar \psi}\psi)\right)\,. 
\label{acteff}
\end{equation}
Model (\ref{acteff}) is equivalent to (\ref{act}) if one
identifies $\beta = \mu\sqrt{\lambda}$,
$\phi\rightarrow\phi/\mu$ and takes the limit
$\mu\rightarrow \infty$. To be more exact, we define the
renormalized Gross-Neveu model as $\mu\rightarrow\infty$
limit of the renormalized effective model (\ref{acteff}),
$\mu$ being the renormalized mass of $\phi$ (see below).
Below we will refer to this identification as the
Gross-Neveu limit. The simplest way to verify the mentioned
equivalence is to consider the path integral representation
for the generating functional:

\begin{equation}
G_{\mbox{\scriptsize eff}}(j) 
= \int\,{\cal D}\phi{\cal D}{\bar\psi}{\cal D}\psi \; 
\exp\left( iS_{\mbox{\scriptsize eff}} +i\int j\phi\right)
\;.
\label{genf}
\end{equation}

Now we perform the Gauss integration over the spinor field
in (\ref{genf}) and then apply the just described
approximation scheme to the resulting functional $G(j)$.
However there is one subtle point about the case $N=1$  in
two dimensions.  As is known, the two-dimensional
Gross-Neveu model at $N=1$ is equivalent to the Thirring
model.  It is known, that the Thirring model does not
reveal  the spontaneous chiral symmetry  breaking, so the
same should be true for the Gross-Neveu model at $N=1$.
The matter is that in the case of  two dimensions and $N=1$
(and {\it only} in this case) the character of the symmetry
of the Gross-Neveu model changes. At the chiral limit
$m\rightarrow 0$ this model is invariant under the discrete
transformation $\psi \rightarrow\gamma_5\psi$. But in the
exclusive case  $N=1$ the symmetry becomes continuous:
$\psi \rightarrow e^{i\alpha\gamma_5}\psi$.  Such an effect
can be easily seen by passing to the spinor components: we
find that non-invariant terms are equal to zero as a
consequence of the grassmanian nature of the spinor field.
So we have the continuous symmetry in the two dimensions,
which cannot be broken spontaneously due to the
Mer\-min-Wag\-ner-Co\-le\-man theorem.

In order to take  into account this fact explicitly  one
could (following, for example, \cite{ZinnJ}) integrate over
only $(N-1)$ components of the spinor field in the
generating functional (\ref{genf}).  The consideration will
be slightly more complicated, but no new effect will arise
except for the multiplier $(N-1)$ instead of $N$ in front
of all the quantities responsible for the chiral symmetry
breaking.  We choose the more transparent way of
integrating over all the components of the spinor field but
should always keep in mind that our consideration is valid
only for $N\ge 2$.

Upon integrating over the spinor field in (\ref{genf}) we
find the following Schwinger-Dyson equation for the
generating functional $G(j)$:

\begin{equation}
(\mu^2 + \partial^2) \vp{G}{j(x)} + N\beta \int d^D\!y\,
{\rm Tr}\left[\Bigl(1-\beta\tr\cdot\vpi{}{j}\Bigr)^{-1}
\!(x,y)\tr(y-x)\right]G-ij(x)G =0\,,
\label{schweq}
\end{equation}
where $\tr(x)$ is the free fermion propagator:

$$
(i\!\not{\!\partial} - m)\tr(x) = -\, \delta(x)\quad,\quad
\tr(x) = \int \frac{d^Dp}{(2\pi)^D}
\frac{\!\not{\!p} + m}{m^2 - p^2}\,,
$$
and $D$ stands for the space-time dimension.

Writing  the generating functional as the sum $G = G^{(0)}
+ G^{(1)} + \dots$,  we put for the leading approximation
$G^{(0)}$: 

\begin{equation}
(\mu^2 + \partial^2) \vp{G^{(0)}}{j(x)} + N\beta \int 
d^D\!y\, {\rm Tr}\left[\Bigl(1-\beta\tr\cdot\vpi{}{j}
\Bigr)^{-1}\!(x,y)\tr(y-x)\right]G^{(0)} = 0\,. 
\label{eqg0}
\end{equation}
The solution to this equation is sought for in the form
$G^{(0)} = \exp(i\sigma*j)$, where $*$ means the space-time
integration and $\sigma$ must be constant by virtue of the
Poincar\'e invariance.  Therefore the Green functions of
the leading approximation are constants and fully
disconnected. Nontrivial connected parts of the Green
functions arise at the further steps of iterations (see
discussion in Section~2).

Now we have from (\ref{eqg0}):

\begin{equation}
i\mu^2\sigma + N\beta\, {\rm Tr}R(0) = 0\,,\label{sig}
\end{equation}
where we adopted the convenient notation $R(x)$:

$$
R(x) \equiv \int d^D\!y  \Bigl(1-\beta\sigma\tr\Bigr)^{-1}
\!(x-y)\tr(y)\equiv (\tr^{-1} - \beta\sigma)^{-1}(x)\,.
$$

The Fourier transformation of the function $R$ reads:

$$
R(x) = \int \frac{d^Dp}{(2\pi)^D}e^{-ipx}\frac{\!\not{\!p}
+ M}{M^2 - p^2}\,,\qquad M\equiv m - \beta\sigma\,.
$$

The first step equation is of the form:

\begin{eqnarray}
\label{firstst}
(\mu^2 + \partial^2) \vp{G^{(1)}}{j(x)} + N\beta \int 
d^D\!y\, {\rm Tr}\left[\Bigl(1-\beta\tr\cdot\vpi{}{j}
\Bigr)^{-1}\!(x,y)\tr(y-x)\right]\hspace*{-3.5mm}
&G^{(1)} =&\\
\hspace*{-3.5mm}& & \hspace*{-3.5mm} ij(x)G^{(0)}\,,
\nonumber
\end{eqnarray}
and we  substitute into this equation the following
expression for the $G^{(1)}$: 

$$ 
G^{(1)} = \Bigl({1\over2}j*{\cal D}*j + i\tau*j\Bigr)
G^{(0)}\,, 
$$ 
where the function ${\cal D}(x)$ and the constant $\tau$
are to be determined from equation (\ref{firstst}).
Omitting some straightforward calculations we write down
the resulting equation:

\begin{eqnarray*}
&&i j(x) = (\mu^2 + \partial^2)\biggl(({\cal D}*j)(x) 
+ i\tau\biggr) - iN\beta^2\int d^D\!y  ({\cal D}*j)(y)
{\rm Tr} \Bigl( R(x-y)R(y-x)\Bigr)\\
&&+ N\beta\,\tau\frac{\partial}{\partial(\beta\sigma)}
{\rm Tr}R(0) - N\beta^3 \int d^D\!y_1 d^D\!y_2  {\cal D}
(y_1 - y_2){\rm Tr} \Bigl( R(x-y_1)R(y_1-y_2)R(y_2-x)
\Bigr) .
\end{eqnarray*}

From the above relation we immediately find the equation
for  $\tau$, which defines the first correction to 
$\sigma$, and the equation for ${\cal D}(x)$, which is of
the main interest for us:

\begin{equation}
(\mu^2 + \partial^2){\cal D}(x) - iN\beta^2 \int d^D\!y 
{\cal D}(x-y){\rm Tr}\Bigl(R(-y)R(y)\Bigr)
 = i\delta(x)\,.\label{eqprop}
\end{equation}

Equations (\ref{sig}) and (\ref{eqprop})  are still formal
due to the ultraviolet divergencies in ${\rm Tr}R(0)$ and
${\rm Tr}( R\cdot R)$ and we should renormalize  our model
in order to give them a definite meaning.

Let us consider now the two-dimensional case  that is put
$D=2$.  Analysis of the divergencies of theory
(\ref{acteff}) shows that it is sufficient to introduce
only one counter term $\delta\mu_{(0)}^2$ to cancel the
divergencies in the equations for $\sigma$ and ${\cal D}$.
Then the leading approximation equation (\ref{sig}) is
modified as follows:

$$
\sigma = -\frac{N\beta}{\mu^2 + \delta\mu_{(0)}^2}\,
\frac{M}{2\pi} \ln\left(1 + \frac{\Lambda^2}{M^2}\right)\,,
$$
where we substitute the value  of the ${\rm Tr}R(0)$
calculated  with a momentum cutoff $\Lambda$.

To verify  the possibility of the spontaneous chiral
symmetry breaking we should study the above equation  at
the chiral limit $m\rightarrow 0$ ($ M \rightarrow
-\beta\sigma$):

\begin{equation}
\sigma = \frac{N\beta^2}{\mu^2 + \delta\mu_{(0)}^2}\,
\frac{\sigma}{2\pi} \ln\left(1 + 
\frac{\Lambda^2}{\beta^2\sigma^2}\right)\,.
\end{equation}
The obvious solution to this equation $\sigma = 0$ leads
(at least up to the few first steps of the scheme) to the
usual perturbative expansion of the generating functional
with massless fermions.  Such a solution is unsatisfactory
one from the physical point of view since it contains the
tachion states which means that the perturbation is carried
over the unstable vacuum \cite{GrNe}.  That is why we will
concentrate on a possible non zero solution, for which we
have:

\begin{equation}
\mu^2 + \delta\mu_{(0)}^2 = \frac{N\beta^2}{2\pi} \ln
\left(1 + \frac{\Lambda^2}{\beta^2\sigma^2}\right)\,.
\end{equation}
Now from equation (\ref{eqprop}) we get the following
expression for the Fourier image ${\cal D}(p^2)$:

$$
{\cal D}(p^2) = \frac{i}{\mu^2 +\delta\mu_{(0)}^2 + 
N\beta^2 \Sigma(p^2) - p^2}\,,
$$
where the bare mass operator $\Sigma$ reads:

\begin{equation}
\Sigma(p^2)\equiv-i \int \frac{d^Dk}{(2\pi)^D}\,{\rm Tr}
R(p+k)R(k)\,.
\label{massoper}
\end{equation}

Let us renormalize ${\cal D}(p^2)$ in the Euclidean
momentum region by the condition:

\begin{equation}
{\rm at }\;\; p^2 = -\omega^2\qquad {\cal D}(-\omega^2) =
\frac{i}{\mu^2}\,.
\label{renorm}
\end{equation}
With the renormalization prescription (\ref{renorm}) we can
find the sum $(\mu^2 + \delta\mu_{(0)}^2)$ and write the
renormalized function ${\cal D}$ and the equation for the
nonzero value of $\sigma$ at the chiral limit as follows:

\begin{equation}
i{\cal D}^{-1}(p^2) = \mu^2  - (p^2 + \omega^2) + 
\frac{N\beta^2}{2\pi}\left(f\biggl(
-\frac{p^2}{\beta^2\sigma^2}\biggr) - 
f\biggl(\frac{\omega^2}{\beta^2\sigma^2}\biggr)\right)
\label{propren}
\end{equation}
\begin{equation}
\frac{N\beta^2}{\pi} + \frac{N\beta^2}{2\pi}
f\biggl(\frac{\omega^2}{\beta^2\sigma^2}\biggr)
 = \mu^2 - \omega^2\,,\label{sigren}
\end{equation}
where the function $f(\theta)$ reads:

$$
f(\theta)\equiv \int_0^1\ln \bigl(1 + \theta x(1-x)
\bigr)dx\,.
$$

In the region $p^2>0$, ${\cal D}(p^2)$ defines the
$s$-channel amplitude of the two fermion scattering,
therefore, as directly follows from (\ref{propren}), the
point $p^2=4\beta^2\sigma^2$ is a two particle threshold
and a fermion with non zero mass $m_F=\beta\sigma$ exists
in our theory.

Now we take the Gross-Neveu limit:
$\beta=\mu\sqrt{\lambda}$, $\mu\rightarrow\infty$.  It can
be shown that at this limit the constant $\sigma$ is
proportional to the fermion condensate :

\begin{equation}
\sigma\;\longrightarrow \; \sqrt{\lambda}\,\frac{\con}{\mu}
\quad \Rightarrow \quad \beta\sigma\; \longrightarrow \;
\lambda\con\,,
\end{equation}
therefore

\begin{equation}
m_F = \lambda\con\,.
\end{equation}

On the other hand the product $(i\beta)^2{\cal D}(p^2)$ in
the Euclidean region $p^2=-q^2<0$ tends to the running
coupling constant of the Gross-Neveu model:

$$
(i\beta)^2{\cal D}(-q^2)\;\longrightarrow\; 
-i\lambda_r\Bigl(\frac{q^2}{\omega^2};\frac{m_F^2}{\omega^2}
\Bigr)\,.
$$
Calculating the limit explicitly we find the running
coupling for the Gross-Neveu model:

\begin{equation}
\lambda_r\Bigl(\frac{q^2}{\omega^2};\frac{m_F^2}{\omega^2}
\Bigr) = \frac{\lambda}{1 + \frac{N\lambda}{2\pi}
\Bigl(f(q^2/m_F^2) - f(\omega^2/m_F^2)\Bigr)}\,.
\label{runcoup}
\end{equation}
It is obvious from (\ref{runcoup}) that  $\lambda =
\lambda_r(1;\frac{m_F^2}{\omega^2})$. Then, denoting the
massless combination $m_F^2/\omega^2$ as $\xi$ we find from
(\ref{sigren}): 

\begin{equation}
\lambda_r^{-1}(1;\xi)= \frac{N}{\pi}\sqrt{1+4\xi}\,
{\rm arcsinh}\frac{1}{\sqrt{4\xi}}  \,,
\label{lrin}
\end{equation}
which gives the final expression for the coupling constant
(\ref{runcoup}):

\begin{equation}
\lambda_r\Bigl(\frac{q^2}{m_F^2}\Bigr) 
=\frac{\pi}{N\sqrt{1+\frac{4m_F^2}{q^2}}\,{\rm arcsinh}
\sqrt{\frac{q^2}{4m_F^2}}}\,.
\label{runcoup2}
\end{equation}

No free parameters are available  except for the fermion
mass $m_F$.  It is in the full agreement with the number of
parameters of the initial Gross-Neveu model: one
dimensionless parameter $\lambda$ is changed for one mass
parameter $m_F$ \cite{GrNe}.

In the deep Euclidean region $q^2\rightarrow\infty$, the
running coupling constant asymptotically vanishes

\begin{equation}
\lambda_r\Bigl(\frac{q^2}{m_F^2}\Bigr)  \sim \frac{\pi}{N
\ln\sqrt{\frac{q^2}{m_F^2}}}\,,
\end{equation}
which reflects the known property of the asymptotical
freedom of the Gross-Neveu model in the two dimensions.

\section{D=3}
\label{d3}

The three dimensional case is interesting in the two
aspects.  Firstly, the four fermion interaction model
(\ref{act}) is not perturbatively renormalizable, since the
mass dimension of the coupling $\lambda$ is negative. But
the effective Yukawa theory (\ref{acteff}) is still
renormalizable (even super renormalizable) model.
Therefore, our approach is a method for handling the
nonrenormalizable theory. And secondly, the dimensional
regularization plays a distinguished role in the three
dimensions, due to a typical singularity of a  one-loop
term  of the theory (\ref{acteff}) has the form
$\Gamma(1-D/2)$ or $\Gamma(2-D/2)$ and, therefore, the
divergencies are absent in this regularization if $D=3$.
The theory is finite and no renormalization is needed. That
is why we make the calculations in two regularizations: the
dimensional and momentum cutoff ones and compare  results.

The form of the main equations (\ref{sig}) and
(\ref{eqprop}) does not depend on the concrete value of the
space-time dimension, so we start our consideration
directly with  those equations.
\vskip 2mm

{\it Dimensional regularization}
\vskip 2mm

On calculating the quantity ${\rm Tr}R(0)$ in the
$D=3-2\varepsilon$ dimensions we obtain from (\ref{sig})
the equation

$$
\mu^2\sigma +3\frac{\Gamma(-{1/2})}{(4\pi)^{3/2}}N\beta 
M^2 = 0\,,
$$
which transforms at the limit $m\rightarrow 0$ to the
following result:

\begin{equation}
\sigma\left(\mu^2 - \frac{3N\beta^3}{4\pi}\sigma\right) 
= 0\,.
\label{sig3}
\end{equation}

We have two solutions again: the trivial $\sigma = 0$ and
the nontrivial $\sigma = \frac{4\pi\mu^2}{3N\beta^3}$ ones.
The nontrivial solution gives the nonzero condensate at the
Gross-Neveu limit:

\begin{equation}
\beta\sigma\;\longrightarrow\;\lambda\con = 
\frac{4\pi}{3N\lambda}\,.
\label{condimreg}
\end{equation}

From equation (\ref{eqprop}) we find

\begin{equation}
i{\cal D}^{-1}(p^2) = \mu^2 - p^2 +\frac{3N\beta^2}{2\pi}
\int_0^1dx \sqrt{\beta^2\sigma^2 - p^2x(1-x)}\,,
\end{equation}
from which in turn we get the mass of a fermion:

\begin{equation}
m_F = \beta\sigma = \lambda\con =  \frac{4\pi}{3N\lambda}
\,.
\end{equation}

In this case the running coupling in the Euclidean region
$p^2 = - q^2 < 0$ reads:

\begin{equation}
\lambda_r^{-1}(q^2) = 
\frac{3Nm_F}{4\pi}\left( 1+ \frac{m_F}{\sqrt{q^2}}
\biggl(1+ \frac{q^2}{4m_F^2}\biggr)
 {\rm arcsin}\sqrt{\frac{q^2}{4m_F^2+q^2}}
\;\right)\,.
\label{runcoup3}
\end{equation}
Its asymptotics in the deep Euclidean region
$q^2\rightarrow \infty$ is 

$$
\lambda_r(q^2) \sim \frac{32}{3N\sqrt{q^2}}\,,
$$
i. e.,  the Gross-Neveu model is asymptotically free in the
three dimensions.

\vskip 2mm

{\it The momentum cutoff  regularization}
\vskip 2mm

The one-fermion loop ${\rm Tr}(R\cdot R)$ and the fermion
tadpole ${\rm Tr}R(0)$ possess the linear divergence, when
calculated at the momentum cutoff $\Lambda$, therefore we
introduce a counter term $\delta\mu_{(0)}^2$  to cancel the
divergence at the leading approximation. For the nonzero
value of $\sigma$ we have the following equation

\begin{equation}
\mu^2 + \delta\mu_{(0)}^2 = \frac{3 N \beta^2}{2\pi^2}
\left(\Lambda + \frac{\pi}{2}\beta\sigma
\right)\,.
\label{sig3cut}
\end{equation}
Equation  (\ref{eqprop}) gives:

$$
i{\cal D}^{-1}(p^2) =  \mu^2 + \delta\mu_{(0)}^2  - p^2 + 
\frac{3N\beta^2}{2\pi}\left(\int_0^1dx\sqrt{\beta^2\sigma^2
-p^2x(1-x)}-\frac{\Lambda}{\pi}\right)\,.
$$

For the renormalization of the theory we take the same
scheme (\ref{renorm}) as we did in the previous section, i.
e., we put

$$
{\cal D}(-\omega^2) = \frac{i}{\mu^2}
$$
at some Euclidean momentum $p^2 = -\omega^2$. We note, that
due to the leading approximation equation (\ref{sig3cut})
the renormalized function   ${\cal D}$ does not depend on
the concrete regularization scheme and as a consequence so
does the running coupling.

Using  the renormalization prescription we can define the
sum $\mu^2+\delta\mu_{(0)}^2$ and get the renormalized
equation for the nonzero constant $\sigma$ from
(\ref{sig3cut}).  At the Gross-Neveu limit the equation can
be written in the form  :

\begin{equation}
\sin^2z + b( z-\sin2z) = 0\;\;,\quad b\equiv \frac{3N}{8\pi}
\lambda\omega\,,
\label{eqcon}
\end{equation}
where we have denoted 

$$
z = {\rm  arcsin}\frac{\omega}{\sqrt{\omega^2 + 
4\lambda^2\con}}\qquad z\in[0,\pi/2]\,,
$$
and have substituted $\beta\sigma \rightarrow \lambda\con$.
One can show that equation (\ref{eqcon})  always has a
nonzero solution $z\in[0,\pi/2]$.  Under the condition
$\omega\ll \lambda\con$ the solution is of the form

$$
\lambda\con = \frac{4\pi}{3N\lambda}\,.
$$

The analytic structure of the renormalized function ${\cal
D}$ allows us to conclude that  there exists a fermion with
the mass $m_F = \beta\sigma$ in our model and to get a
running coupling constant in the momentum cutoff
regularization:

$$
\lambda_r^{-1}(q^2) = 
\frac{3Nm_F}{4\pi}\left( 1+ \frac{m_F}{\sqrt{q^2}}
\biggl(1+ \frac{q^2}{4m_F^2}\biggr)
{\rm arcsin}\sqrt{\frac{q^2}{4m_F^2+q^2}}
\;\right)\,,
$$
which is identical to the result of the dimensional
regularization (\ref{runcoup3}).

The only parameter of the model is the fermion mass, which
defines the strength of interaction at the classical limit
$q\rightarrow0$:

\begin{equation}
\lambda_r(0)  = \frac{4\pi}{9Nm_F}\,.
\end{equation}
The similar results were obtained in the framework of $1/N$
expansion
\cite{Rosen}.

\section{D=4}
\label{d4}

The case of the four dimensional space-time is more
involved due to the effective theory contains the
divergencies in the four point Green function of the scalar
field. Therefore we have to introduce into the action the
corresponding self-interaction of the scalar field $\sim
\phi^4$. Thus we start from the action

\begin{equation}
S_{\mbox{\scriptsize eff}} =  \int\,d^4x 
\left({\bar \psi}\,(i\!\not{\!\partial} - m)\,\psi
-{1\over 2}\phi(Z_{\phi}\partial^2 + Z_{\mu}\mu^2)\phi +
\beta\,\phi({\bar \psi}\psi) - \frac{Z_4\beta^4}{4!}
\phi^4\right)\,.
\label{acteff4}
\end{equation}
Here the  constants $Z_a $ represent the boson field
renormalization whereas the fermion renormalization
multipliers are considered as being absorbed into the norm
of the fermion fields and into the constants of the action.

The question of equivalence of the model (\ref{acteff4})
(also referred to as "extended Gross-Neveu model") to the
Gross Neveu model (\ref{act}) becomes nontrivial in the
case $D=4$. It is evident that such an equivalence can take
place in the nonperturbative sense only. This equivalence
(in chiral limit) was motivated in the framework of
$1/N$-expansion (see, for example \cite{ZinnJ}) and is
based on the fact that in this limit the terms
$\phi\,\partial^2\phi$ and $\phi^4$ of effective action
(\ref{acteff4}) are irrelevant when one defines physical
quantities in the critical infrared region. As will be seen
below the same arguing can be applied in our approach too.
Therefore we {\it define \/} the Gross-Neveu model in $D=4$
as the Gross-Neveu limit of the renormalized model
(\ref{acteff4})

We will use the dimensional regularization in this section.
As the leading approximation to the Schwinger-Dyson
equation of the model (\ref{acteff4}) we have the
following:

\begin{eqnarray}
(Z^{(0)}_{\phi}\partial^2 + Z^{(0)}_{\mu}\mu^2)
\hspace*{-5mm}&&\hspace*{-1mm}\vp{G^{(0)}}{j(x)}
-\frac{Z^{(0)}_4\beta^4}{3!}\vp{^3G^{(0)}}{j^3(x)} 
+\nonumber\\
&&\hspace*{-2mm} N\beta \int d^4\!y\, {\rm Tr}\left[\Bigl(
1-\beta\tr\cdot\vpi{}{j}\Bigr)^{-1}\!(x,y)\tr(y-x)\right]
G^{(0)} = 0
\end{eqnarray}
The solution to this equation is sought for in the form
$G^{(0)}=\exp(i\sigma*j)$, which gives the connection of
the leading counter terms:

\begin{equation}
Z^{(0)}_{\mu}\mu^2\sigma + \frac{Z^{(0)}_4}{3!}\beta^4
\sigma^3 = iN\beta{\rm Tr}R(0)\,.
\label{sig4}
\end{equation}

The principal difference with the previous cases $D=2,3$
is that we have {\it two } counter terms in (\ref{sig4})
and upon their fixing equation (\ref{sig4}) turns into an
identity and does not define any specific value of
$\sigma$.  We will seek for the first step approximant in
the same form as in the previous sections and find the
following result for the Fourier image ${\cal D}(p^2)$:

\begin{equation}
i{\cal D}^{-1}(p^2) =  Z^{(0)}_{\mu}\mu^2+ Z^{(0)}_4
\frac{\beta^4\sigma^2}{2}+N\beta^2\Sigma(p^2) - 
Z^{(0)}_{\phi}p^2\,,
\label{FourD}
\end{equation}
where the mass operator $\Sigma$ is one-fermion loop
(\ref{massoper}).  Choose the renormalization prescription
for  ${\cal D}$ at an Euclidean point $p^2=-\omega^2$ :

$$
{\rm at} \quad p^2\simeq -\omega^2 \qquad 
i{\cal D}^{-1}(p^2) = \mu^2 + O\bigl((p^2+\omega^2)^2\bigr)
\,.
$$
This prescription fixes the value of the counter terms
$Z^{(0)}_{\phi}$ and the sum $Z^{(0)}_{\mu}\mu^2 +
Z^{(0)}_4\beta^4\sigma^2/2$. Taking into account equation
(\ref{sig4}) allows one to define all the leading
approximation counter terms.

Let us consider the chiral limit $m\rightarrow 0$. First of
all it is worth mentioning that equations (\ref{sig4}) and
(\ref{FourD}) up to notations coincide with analogous
equations of $1/N$-expansion method.  Therefore the above
mentioned arguments of \cite{ZinnJ} about the equivalence
are valid for our consideration too.

We note now that there exist two different  cases which do
not contradict our equations: the first of them is
$\sigma=0$ and the second one is $\sigma\not=0$.  Let us
begin with the case of the nonzero $\sigma$ and take for
simplicity $\omega = 0$. The  renormalized function ${\cal
D}$ has the form

\begin{equation}
i{\cal D}^{-1}(p^2) = \mu^2 - \frac{N\beta^2}{8\pi^2}
\Biggl(p^2 + 6 m_F^2\Phi\Bigl(-\frac{p^2}{m_F^2}\Bigr)
\Biggr)\,,
\label{prop4}
\end{equation}
where the function $\Phi(\theta)$ stands for the integral:

$$
\Phi(\theta) = \int_0^1
\bigl(1+\theta x(1-x)\bigr)\ln\bigl(1+\theta x(1-x)\bigr)
dx\,
$$
and $m_F^2=\beta^2\sigma^2$.  On putting as in the above
sections 
$$ 
(i\beta)^2{\cal D}(-q^2)\; \longrightarrow\;  -
i \beta_r(q^2) 
$$ 
we obtain from  (\ref{prop4}) the expression for the Yukawa
running coupling $\beta_r(q^2)$

\begin{equation}
\beta_r(q^2) = \frac{\beta_r(0)}{1 + 
\frac{N\beta_r(0)}{8\pi^2}\left(q^2 - 6 m_F^2\Phi\bigl(
\frac{q^2}{m_F^2}\bigr)
\right)}\,.\label{run4}
\end{equation}
Since equations (\ref{sig4}) are fulfilled identically, the
parameter $\beta_r(0)$ is as free as the mass $m_F$ is.

Now let us investigate the behaviour of the denominator of
expression (\ref{run4}) in the regions of small and large
values of $q^2$. It can be easily seen that at
$q^2\rightarrow 0$ this denominator tends to the unity
while in the deep Euclidean region $q^2\rightarrow \infty$
its asymptotics is $-q^2\ln q^2\rightarrow -\infty$. Due to
the denominator being a continuous function there exists
{\it a finite} value $q_0^2$ at which the running coupling
constant has a pole.  This indicates to the presence of a
tachion state in our model. The only way to save the
situation is to put the free parameter $\beta_r(0)$  be
equal to zero.

The same is true for another phase of the model, when
$\sigma=0$.  The only difference is that we must keep an
arbitrary nonzero value of the renormalization point
$\omega$  in order to avoid the infrared singularities at
the chiral limit.  The running  coupling reads:

\begin{equation}
\beta_r(q^2) = \frac{\beta_r(\omega^2)}{1 + 
\frac{N\omega^2\beta_r(\omega^2)}{8\pi^2}F(q^2/\omega^2)}
\,,
\end{equation}
where $F(x) = x(1-\ln x) - 1$. The function  $F(x)$ takes
its maximum at the point $x=1$: $F_{max}=0$ and is negative
for all other positive $x$.  Therefore we come to the same
conclusion as above: 

{\bf (i)}
if $\frac{8\pi^2}{N}\le\omega^2\beta_r(\omega^2)$, two
tachion states are present and the model is contradictory.

{\bf (ii)}
if $0< \omega^2\beta_r(\omega^2)<\frac{8\pi^2}{N}$, one
tachion state is present and the model is contradictory
again.

{\bf (iii)}
if $\beta_r(\omega^2)  = 0$, the tachion is absent but the 
running coupling is identical zero. 

From the analysis above it follows that the renormalized
effective coupling constant of Yukawa interaction
$\phi{\bar \psi}\psi$ should be zero.  This means that the
renormalized model (\ref{acteff4}) describes a free fermion
and a scalar boson with selfinteraction $\phi^4$ decoupled
from each other due to trivialization of Yukawa coupling.
Further, the triviality of the theory $\phi^4$ in $D=4$ is
well known and was demonstrated in various nonperturbative
approaches (see for example \cite{Roch1} and references
therein). It is worth pointing out that this triviality can
be shown by means of the same method we are using in the
present work \cite{Roch1}. In the Gross-Neveu limit
$\mu\rightarrow\infty$ the free boson disappears from the
spectrum and we come to the theory of free fermion.

Note, that the triviality of four-fermion interaction in
$D=4$ was also found in the framework of $1/N$-expansion
and confirmed by lattice simulations \cite{Kim}. Our
calculation is one more argument in favor of such a
trivialization.

\section{The bilocal source}
\label{biloc}

In this section we would like to discuss another approach
to the Gross-Neveu model. The approach is based on the
bilocal source formalism, which is more natural and
informative than that of the preceding sections. We will
directly deal with the fermion degrees of freedom without
integrating over them.  This allows one to get the full
information on the fermion dynamics.  Namely it is possible
to calculate the fermion Green functions like the
propagator, the amplitude and so on.  Besides, this method
is  technically  simpler.  But as a price for the above
advantages the use of the bilocal source encounters a
problem, which makes the results obtained in this formalism
to be not very firm from the position of the mathematical
rigour. The problem originates from the well known fact
that keeping the right  Bose or Fermi statistical
properties of the Green functions is a rather nontrivial
task, when the bilocal source is used. We postpone the
discussion of this difficulty to the end of the section.

Now let us demonstrate how the results of the previous
sections can be obtained in the bilocal source formalism.
We consider the two dimensional $N$-component Gross-Neveu
model with the action $S_{GN}$ (\ref{act}) and introduce a
bilocal source --- the function $\eta_{\alpha\beta}(x,y)$
depending on two space-time points and two multi indices  
$\alpha$ and $\beta$ including the color and Lorenz degrees
of freedom. The $n$-th derivative of the generating
functional

$$
G(\eta) = \int {\cal D}\bar \psi{\cal D}\psi \exp(iS_{GN} 
- i\bar \psi*\eta*\psi)
$$
defines the $2n$-th point Green function. This is an
advantage of the bilocal source, since to find a Green
function one has to calculate half as many derivatives than
when a local source is used. The Schwinger-Dyson equation
for the functional $G(\eta)$ is of the form:

\begin{eqnarray}
\label{sch}
i\lambda\vp{^2G}{\eta_{\beta\alpha}(yx)\delta
\eta_{\gamma\gamma}(xx)}
+ (i\imb{\partial}_{x} -m)_{\alpha\gamma}\vp{G}{
\eta_{\beta\gamma}(yx)}
+ \delta_{\alpha\beta}\delta(x\hspace*{-3mm}&-&
\hspace*{-3mm}y)G \\ 
&=& \int d^2z\eta_{\alpha\gamma}(xz)\vp{G}{\eta_{
\beta\gamma}(yz)}\,,\nonumber
\end{eqnarray}
where the summation over the repeated indices is assumed.
The leading approximation $G^{(0)}$ obeys (\ref{sch}) with
zero right site. Substituting $G^{(0)} = \exp\{{\rm
Tr}(\eta*\tr)\}$, we find the equation for the leading
approximation to the fermion propagator $\tr$:

\begin{equation}
(M - i\imb{\partial})\tr(x) = \delta(x)\,,
\label{propag}
\end{equation}
where the mass parameter $M$ denotes the combination:

\begin{equation}
M = m - i\lambda\,{\rm Tr}\tr(0)\label{consist}\,.
\end{equation}
Since ${\rm Tr}\tr(0)$ is a function of M, relation
(\ref{consist}) is a consistency condition, which is an
analog of the gap equation.

To solve the next step equation 

\begin{equation}
i\lambda\vp{^2G^{(1)}}{\eta\delta\eta}
+ (i\imb{\partial} -m)\vp{G^{(1)}}{\eta}
+ G^{(1)} = \eta*\vp{G^{(0)}}{\eta}\,,
\label{1st}
\end{equation}
we substitute the expression

\begin{equation}
G^{(1)} = \Bigl(\frac{1}{2} {\rm Tr}_{(12)}F_{12}*\eta_1*
\eta_2 + {\rm Tr}\tr^{(1)}*\eta\Bigr)G^{(0)}\label{g1}\,,
\end{equation}
where the function
$F^{\alpha_1\alpha_2}_{\beta_1\beta_2}(x_1x_2|y_1y_2)$ is
connected with the leading approximant for the two-fermion
amplitude and $\tr^{(1)}$ is the first correction to the
propagator.  The equations for $F$ and $\tr^{(1)}$ can be
easily obtained from (\ref{1st})  and (\ref{g1}), so we
directly write down the answer for the amplitude omitting
the explicit form of the mentioned equations:

\begin{eqnarray}
&&\hspace*{-10mm}F^{\alpha_1\alpha_2}_{\beta_1\beta_2}
(x_1x_2|y_1y_2) = \nonumber\\
&&\hspace*{-10mm}\int dz_1dz_2 \Bigl(\tr\,(x_1-z_1)\cdot
\tr\,(z_1-y_1)\Bigr)_{\alpha_1\beta_1}\,{\cal K}(z_1 - z_2)
\,\Bigl(\tr\,(x_2-z_2)\cdot \tr\,(z_2-y_2)
\Bigr)_{\alpha_2\beta_2}\\
&&\hspace*{-10mm} + \;\dots\nonumber\,,
\end{eqnarray}
where the dots stand for the disconnected part.

The scalar kernel $\cal K$ is a solution of the equation

\begin{equation}
\int dy (1 - i\lambda \Sigma)(x-y){\cal K}(y) = -i\lambda
\delta(x)
\label{k}
\end{equation}
and the function $\Sigma(x)$ is a fermion loop (compare
with Section~3)

$$
\Sigma(x) = {\rm Tr}\tr(x)\tr(-x) \,.
$$

Equation (\ref{k}) can be easily solved in the momentum
space and the Fourier image of the bare kernel ${\cal K}$
turns out to be:

\begin{equation}
{\cal K}(p^2) = - \frac{i\lambda}{1 - i\lambda\Sigma(p^2)}
\,.
\end{equation}

Let us consider the chiral limit $m\rightarrow 0$ of our
model.  To renormalize the amplitude at this stage we
should introduce only one  counter term $\delta
\lambda_{(0)}$.  We require the connected part of the
amplitude at the symmetric point in the Mandelstam
variables to be:

\begin{equation}
F^{\mbox{\scriptsize conn}}(s=t=-\omega^2)\equiv {\cal K}
(-\omega^2) = -i\lambda_{r}\label{rensch4}\,,
\end{equation}
where $s$ and $t$ are the Mandelstam variables. This allows
us to fix the counter term $\delta\lambda_{(0)}$ and to
obtain the renormalized connected amplitude

\begin{equation}
{\cal K}^{\mbox{\scriptsize ren}}(p^2) = - 
\frac{i\lambda_{r}}{1 + i\lambda_{r}(\Sigma(-\omega^2)
- \Sigma(p^2))}\,.
\label{renamp}
\end{equation}

Now we should go back to  the consistency condition
(\ref{consist}) which takes the following form at the
chiral limit $m\rightarrow 0$

$$
M =  - i(\lambda + \delta\lambda_{(0)})\,{\rm Tr}\tr(0)\,.
$$
Taking into account the value of $\delta\lambda_{(0)}$
fixed by the renormalization scheme (\ref{rensch4}) we find
an equation defining possible values of the fermion mass in
our model. Besides the trivial solution $M=0$, there exists
the nontrivial one $M=m_F\not=0$ which corresponds to the
spontaneous symmetry breaking and gives a connection among
the fermion physical mass $m_F$, the renormalized coupling
$\lambda_r$ and the subtraction point $\omega^2$ (compare
with (\ref{lrin}):

\begin{equation}
N\sqrt{1+\frac{4m_F^2}{\omega^2}}\;{\rm arcsinh}\sqrt{
\frac{\omega^2}{4m_F^2}} = \frac{\pi}{\lambda_r}\,.
\label{connection}
\end{equation}
With the help of (\ref{connection}) we can get  the running
coupling constant from (\ref{renamp})

\begin{equation}
\lambda_r\Bigl(\frac{q^2}{m_F^2}\Bigr) 
=\frac{\pi}{N\sqrt{1+\frac{4m_F^2}{q^2}}\,{\rm arcsinh}
\sqrt{\frac{q^2}{4m_F^2}}}\,,
\end{equation}
which is identical to (\ref{runcoup2}).

Turning to the discussion at the beginning of this section,
where is a vulnerable place of the  above consideration? It
is contained in the first term of the Schwinger-Dyson
equation (\ref{sch}). Indeed it can be easily shown that
due to the Fermi statistics of the spinor fields the
generating functional $G(\eta)$ obeys identically the
relation:

\begin{equation}
\vp{^2G}{\eta_{\beta_1\alpha_1}(y_1x_1)\delta\eta_{\beta_2
\alpha_2}(y_2x_2)} = - \vp{^2G}{\eta_{\beta_1\alpha_2}
(y_1x_2)\delta\eta_{\beta_2\alpha_1}(y_2x_1)}\,.
\label{fermi}
\end{equation}
Therefore we can write the first term of (\ref{sch}) as

\begin{equation}
- \vp{^2G}{\eta_{\beta\gamma}(yx)
\delta\eta_{\gamma\alpha}(xx)}\,\label{subs}
\end{equation}
and the equation thus obtained will be equivalent to
(\ref{sch}) from the point of view of the {\it full}
generating functional which is a strict solution of the
Schwinger-Dyson equation. But for our approximant $G^{(0)}$
such a change has a crucial consequence since it does not
possess the property (\ref{fermi}). And the substitution
(\ref{subs}) leads to the appearance of the term 
$\tr_{\alpha\gamma}(x) \tr_{\gamma\beta}(0)$ instead of
$\tr_{\alpha\beta}(x){\rm Tr}\tr(0)$ in  equation
(\ref{propag}). But such an equation for $\tr(x)$ does not
possess any solution except for the perturbative one.  The
possible  way to resolve this contradiction is in the use
of properly symmetrized form of the first term in equation
(\ref{sch}). Such a new equation will be still equivalent
to (\ref{sch}) for the full generating functional and will
respect the property (\ref{fermi}), when the approximate
solutions is used.

It can also  be shown that the important property
(\ref{fermi}) which is responsible, in particular, for the
crossing symmetry is restored for the Green functions
successively step by step of the approximation scheme. For
example, all the Green functions calculated from the
$G^{(0)}$ will consist of disconnected parts only and will
not possess the true structure demanded by the Fermi
statistics. But taking into account the first correction
$G^{(1)}$ not only gives the leading approximation to the
connected part of the amplitude but also restores the true
statistical structure for the disconnected part of the four
point Green function approximant. At the next steps we will
successively find corrections improving the connected and
disconnected parts of the higher Green functions. The use
of the symmetrized form of the Schwinger-Dyson equation
allows one to get the properly symmetrized Green functions
directly at the corresponding steps of the scheme.

\section*{Acknowledgements}
This work is  supported in part by the Russian Foundation
of Fundamental Researches, grant 95-02-03704, and the work
of P.A.S. is supported in part by the grant INTAS-RFBR
95-0829.

\end{document}